\documentclass[letterpaper]{article} 
\usepackage{aaai25}  
\usepackage{times}  
\usepackage{multirow}
\usepackage{helvet}  
\usepackage{courier}  
\usepackage[hyphens]{url}  
\usepackage{graphicx} 
\usepackage{natbib}  
\usepackage{caption} 
\usepackage{booktabs}
\usepackage{float}
\usepackage{algorithm}
\usepackage{algorithmic}
\usepackage{amsmath}

\usepackage{newfloat}
\usepackage{listings}
\DeclareCaptionStyle{ruled}{labelfont=normalfont,labelsep=colon,strut=off} 
\floatstyle{ruled}
\newfloat{listing}{tb}{lst}{}
\floatname{listing}{Listing}

\usepackage{xcolor}

\newcommand{\model}{\texttt{text2midi}}

\title{Text2midi: Generating Symbolic Music from Captions}
\author {
    Keshav Bhandari\textsuperscript{\rm 1},
    Abhinaba Roy\textsuperscript{\rm 2},
    Kyra Wang\textsuperscript{\rm 2},
    Geeta Puri\textsuperscript{\rm 2},
    Simon Colton\textsuperscript{\rm 1},
    Dorien Herremans\textsuperscript{\rm 2}
}
\affiliations {
    \textsuperscript{\rm 1}Queen Mary University of London\\
    \textsuperscript{\rm 2}Singapore University of Technology and Design\\
    k.bhandari@qmul.ac.uk, abhinaba\_roy@sutd.edu.sg, kyra\_wang@mymail.sutd.edu.sg
}

\usepackage{bibentry}

\begin{document}

\maketitle

\begin{abstract}
This paper introduces \model{}, an end-to-end model to generate MIDI files from textual descriptions. Leveraging the growing popularity of multimodal generative approaches, \model{} capitalizes on the extensive availability of textual data and the success of large language models (LLMs). Our end-to-end system harnesses the power of LLMs to generate symbolic music in the form of MIDI files. Specifically, we utilize a pretrained LLM encoder to process captions, which then condition an autoregressive transformer decoder to produce MIDI sequences that accurately reflect the provided descriptions. This intuitive and user-friendly method significantly streamlines the music creation process by allowing users to generate music pieces using text prompts. We conduct comprehensive empirical evaluations, incorporating both automated and human studies, that show our model generates MIDI files of high quality that are indeed controllable by text captions that may include music theory terms such as chords, keys, and tempo. We release the code and music samples on our demo page\footnote{\url{https://github.com/AMAAI-Lab/Text2midi}} for users to interact with \model{}.
\end{abstract}

%

\section{Introduction}

The rapid advancements in large language models (LLMs) have revolutionized the way we interact with and leverage various forms of media, including text, images, and audio. Researchers have been able to build systems that demonstrate remarkable capabilities in both analyzing diverse input texts as well as generate highly precise text, image, video, as well as audio output \cite{touvron2023llama,ramesh2022hierarchical, ramesh2021zero,melechovsky2023mustango} This has paved the way for a wide array of downstream applications such as question-answering systems \cite{touvron2023llama}, image generation tools \cite{ramesh2021zero}, and even music generation platforms \cite{sunoai}. These breakthroughs have ushered in a new era of multimodal intelligence, where the integration of multiple data sources and modalities can be harnessed to enhance our understanding, creativity, and problem-solving abilities across a wide range of domains. While prior work has explored generating audio music from text, the potential of large language models has not yet been fully realized for symbolic music generation. In this paper, we introduce \model{}, a system that leverages the power of LLMs to generate symbolic music in the form of MIDI files from textual descriptions. 

In Music Information Retrieval (MIR), MIDI is a crucial format that provides a symbolic representation of music \cite{schedl2014music,herremans2017functional}. Its symbolic nature has established it as a popular format for music creation, which is used by music producers and composers in popular Digital Audio Workstations (DAWs).

\begin{figure}[t]
\includegraphics[width=8cm]{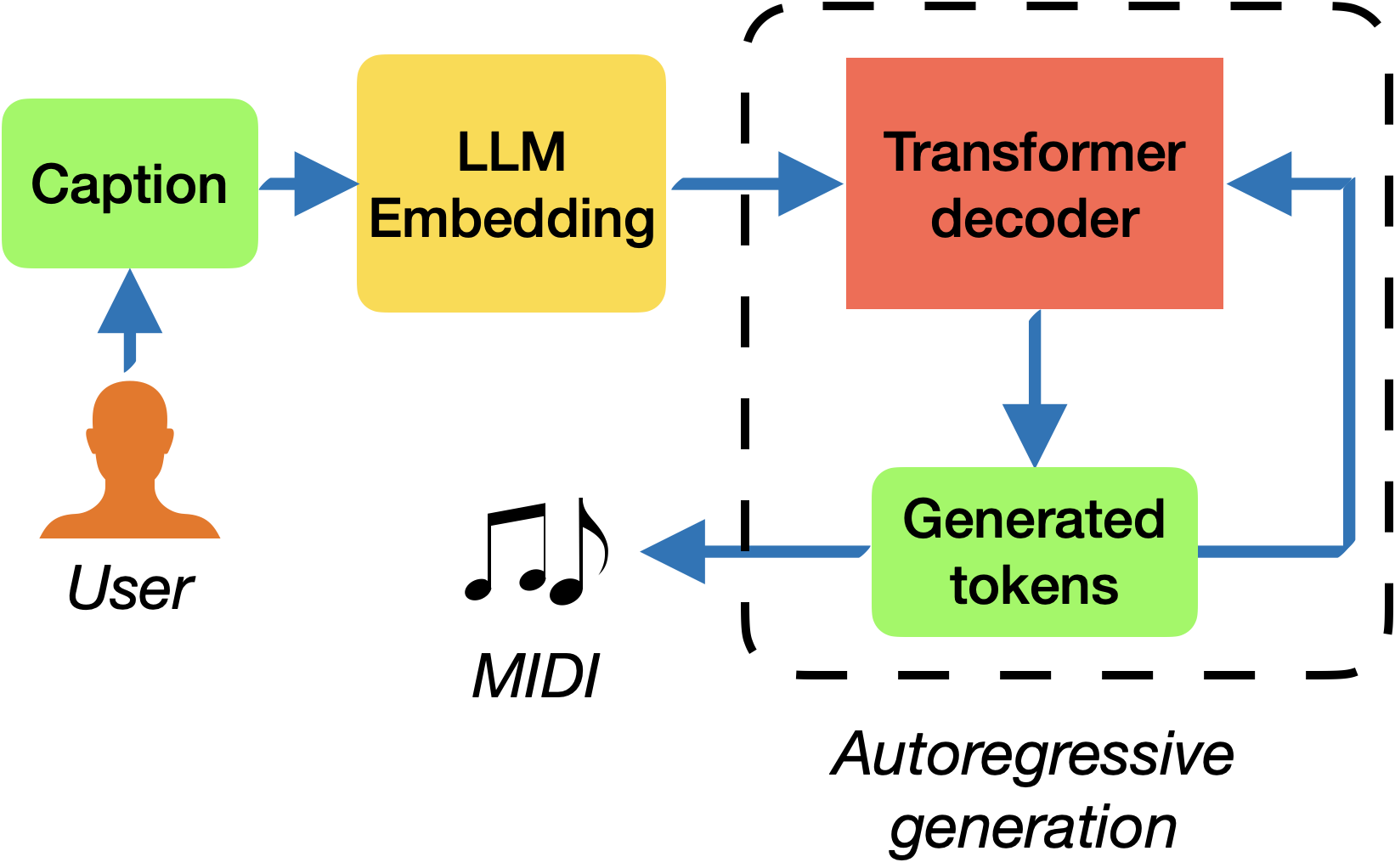}
\caption{Overview of our approach. We use a pretrained LLM encoder to encode input captions. This is passed to the trained transformer decoder which generates encoded MIDI tokens autoregressively.}
\label{fig:overview}
\end{figure}

In recent times, there is a growing emphasis on the creation of music based on unstructured text instructions \cite{copet2023simple, melechovsky2023mustango, huang2023noise2music}. These models leverage large language models to transform textual descriptions of musical attributes into concrete audio compositions. This process requires a precise alignment between textual content and musical features to ensure that the resulting music accurately reflects the provided text guidelines. However, the majority of existing methodologies primarily focus on directly producing \emph{audio} music. In contrast, symbolic music representation, particularly in MIDI format, has not received sufficient attention from researchers within the community. The lack of text-to-midi models can largely be attributed to the lack of a large-scale dataset of MIDI files annotated with text captions. Recently, \cite{melechovsky2024midicaps} expanded upon the Lakh MIDI dataset—which includes 168,407 MIDI files—by incorporating free-form text captions that provide musical insights; referred to as MidiCaps. To date, the only significant contribution comes from \cite{lu2023musecocogeneratingsymbolicmusic}, which adopts an indirect methodology by deriving musical features from textual captions for use as conditions. This approach involves two distinct stages: initially focusing on understanding and extracting musical attribute values from plain text during a comprehension phase; subsequent to this extraction process, these attributes are fed into a generator network designed to produce symbolic music in a second pass. It should be noted that both training and inference phases associated with such a dual-stage process can prove time-consuming.

In this work, we propose \model{}, the first end-to-end model for generating high quality MIDI files directly from text (see Figure \ref{fig:overview}). Our goal is to create an intuitive and accessible solution that caters to both technical and non-technical musicians alike. We employ a transformer architecture based on an encoder-decoder framework capable of processing any unstructured text (caption) as its input. This process involves utilizing a pretrained large language model (LLM) encoder to embed the provided caption, which subsequently acts as conditional data for the decoder to generate MIDI sequences in an autoregressive manner. To maximize our architecture's potential, we implement semi-supervised pretraining using the SymphonyNet dataset \cite{liu2022symphony} containing 46,359 MIDI files. During data preprocessing, we create pseudo captions that encapsulate the musical attributes present in the MIDI files and use the pseudo caption-MIDI pair for pretraining. Subsequently, we leverage Midicaps—a specialized dataset aimed at facilitating text-to-MIDI generation to fine-tune our model and enhance performance. To assess the effectiveness of our training methodology, we conduct subjective human evaluations to gauge the overall musical quality of the MIDI output. Additionally, we also perform objective assessments to verify alignment between musical attributes specified in input captions and those present in the resultant MIDI files. We expect that our efforts will assist both laymen, musicians, and music producers alike, as they can express their concepts through free-flowing text and utilize the generated MIDI files as a foundational element for their compositions, or just as they are. Furthermore, we are confident that our work will motivate researchers within the Music Information Retrieval (MIR) domain to explore MIDI generation tasks more thoroughly.

The main contributions of our work are as follows:
\begin{itemize}
    \item We present \textbf{\model{}}, the first end-to-end model for the task of generating MIDI files from text captions. Our model leverages the power of pretrained large language models to encode the provided textual descriptions and then utilizes an autoregressive transformer decoder to generate the corresponding MIDI sequences.
    \item We utilize a semi-supervised pretraining approach using symphonynet \cite{liu2022symphony} - a dataset with complex multi-track and multi-instrument MIDI files. We first extract relevant musical attributes from this dataset, such as instruments, tempo, and time signature, and generate pseudo captions from these attributes to create caption-MIDI pairs for pretraining our model.   
    \item We are the first to train on the MidiCaps dataset, a curated large-scale collection of MIDI-caption pairs, for the task of text-to-MIDI generation. This dataset allows us to further fine-tune and evaluate our model's performance in generating MIDI compositions from textual descriptions. 
\end{itemize}

\section{Related Work}

The first generative symbolic music model, developed by \citet{10.1162/lmj_a_01037}, composed the famous string quartet `the Illiac Suite' using rules and Markov chains. The field of automatic composition has slowly grown since then, gaining new popularity with the introduction of various types of deep networks such as recurrent neural networks (RNNs), which researchers quickly discovered helped to maintain long-term structure in their models \cite{chuan2018modeling}. For instance, BachBot \cite{liang2016bachbot} and DeepBach \cite{pmlr-v70-hadjeres17a} are two of the popular models based on long-short term memory models (LSTMs) that generate music sequences in the style of Bach. With the introduction of the Transformer as well as steadily growing MIDI datasets, more powerful models such as Music Transformer \cite{huang2018musictransformer} and the recent Museformer model \cite{NEURIPS2022_092c2d45} were developed. The latter leverages the fine-grained and coarse-grained attention in Transformers \citep{vaswani2017attention} to generate fairly long and high quality music sequences. For a more complete overview of this evolution and currently remaining challenges, the reader is referred to \citet{herremans2017functional} and \citet{ji2023survey}.

To make generative music models really  useful, they should be controllable by the user. This way, they provide a way to augment the human composer rather than just replace them. In the literature, we find models that allow the generated MIDI files to be controlled based on an input emotion \cite{makris2021generating}, chords \cite{zixun2021hierarchical}, lyrics \cite{yu2021conditional}, tension \cite{herremans2017morpheus, guo2020variational}, among others. Typically, these models work by feeding an input label with the desired condition, which are processed through cross attention in transformer networks, or other types of conditional neural networks. 

In a recent evolution of the field of generative audio, a number of text-to-music models have been developed in the last two years \cite{melechovsky2023mustango, copet2023simple, huang2023noise2music}. These models generate audio directly based on a free-form text prompt. They typically leverage pretrained large language text encoders, such as FLAN-T5 \cite{chung2024scaling}, and feed the resulting embedded text representation as input to the audio decoder. Some models generate shorter fragments using diffusion \cite{melechovsky2023mustango}, whereas others allow longer fragments through an autoregressive transformer approach \cite{copet2023simple}. In this paper, we follow a similar approach as the latter, except on symbolic music.   

In the field of generative MIDI models, we have not yet seen such an evolution. The only text to MIDI model currently is MuseCoco \cite{lu2023musecocogeneratingsymbolicmusic}. MuseCoco generates symbolic music from text descriptions by leveraging a two-stage framework that constitutes text-to-attribute understanding and attribute-to-music generation. It provides more control over music elements which has always been a pain point for previous models. The recently released chatGPT4o\footnote{\url{chatgpt.com}} is also able to generate MIDI files based on a text prompt, however, after a quick examinations, the files are mostly nonsensical or barely contain notes. A similar conclusion was drawn by \citet{lu2023musecocogeneratingsymbolicmusic}. 
Another attempt at developing symbolic music from text captions is due to \cite{zhang2020butter}, who developed the BUTTER (Better Understanding of Text To Explanation and Recommendation) model. This model generates music in ABC format \footnote{\url{https://abcnotation.com/}}based on a text caption. Like MuseCoco, the model also takes a two-stage approach and predicts key words such as music attributes from the text caption to generate folk music.  
Finally, \citet{wu2022exploring} explores how pretrained language models can be used to generated ABC format music. However, the ABC format is limited in terms of representing polyphonic multitrack music. 

In this work, we aim to take a holistic approach and train an end-to-end model directly on the recently released MidiCaps dataset \cite{melechovsky2024midicaps}. This dataset contains 168,385 MIDI files that have been annotated with rich text captions. In the next section we discuss out model architecture, which is followed by results of our experiments and discussion.

\section{Method}
Here, we first discuss the mathematical formulation for our problem followed by the transformer architecture used and the semi-supervised pretraining. 

\subsection{Mathematical Formulation}
The input text \(T\) is first encoded using a pretrained Flan T5 encoder \(E\), as \(H_T = E(T)\), where \(H_T\) is an \(n \times d\) dimensional vector, \(n\) is the token or text sequence length, and \(d\) is the T5 feature dimension. MIDI data \(M\) is tokenized using the REMI tokenizer to generate the sequence \(S_M = M_{\text{tok}}(M)\). The decoder \(D(H_T, S_M)\) utilizes the sequence \(S_M\) and \(H_T\) as hidden states to predict the next MIDI token at time step \(m\):
\[
P(S_{m+1} \mid H_T, S_m) = \text{softmax}(D(H_T, S_m))
\]
where \(S_m\) represents MIDI tokens up to time step \(m\). Finally, a REMI decoder \(M_{\text{detok}}\) converts the generated tokens back to MIDI:
\[
M_{\text{gen}} = M_{\text{detok}}(D(E(T), S_M))
\]

\subsection{Architecture}

\subsubsection{Transformer conditioning with Text}

\begin{figure}[h!]
\centering
\includegraphics[trim=2.5cm 0.8cm 0 1cm, clip, width=8cm]{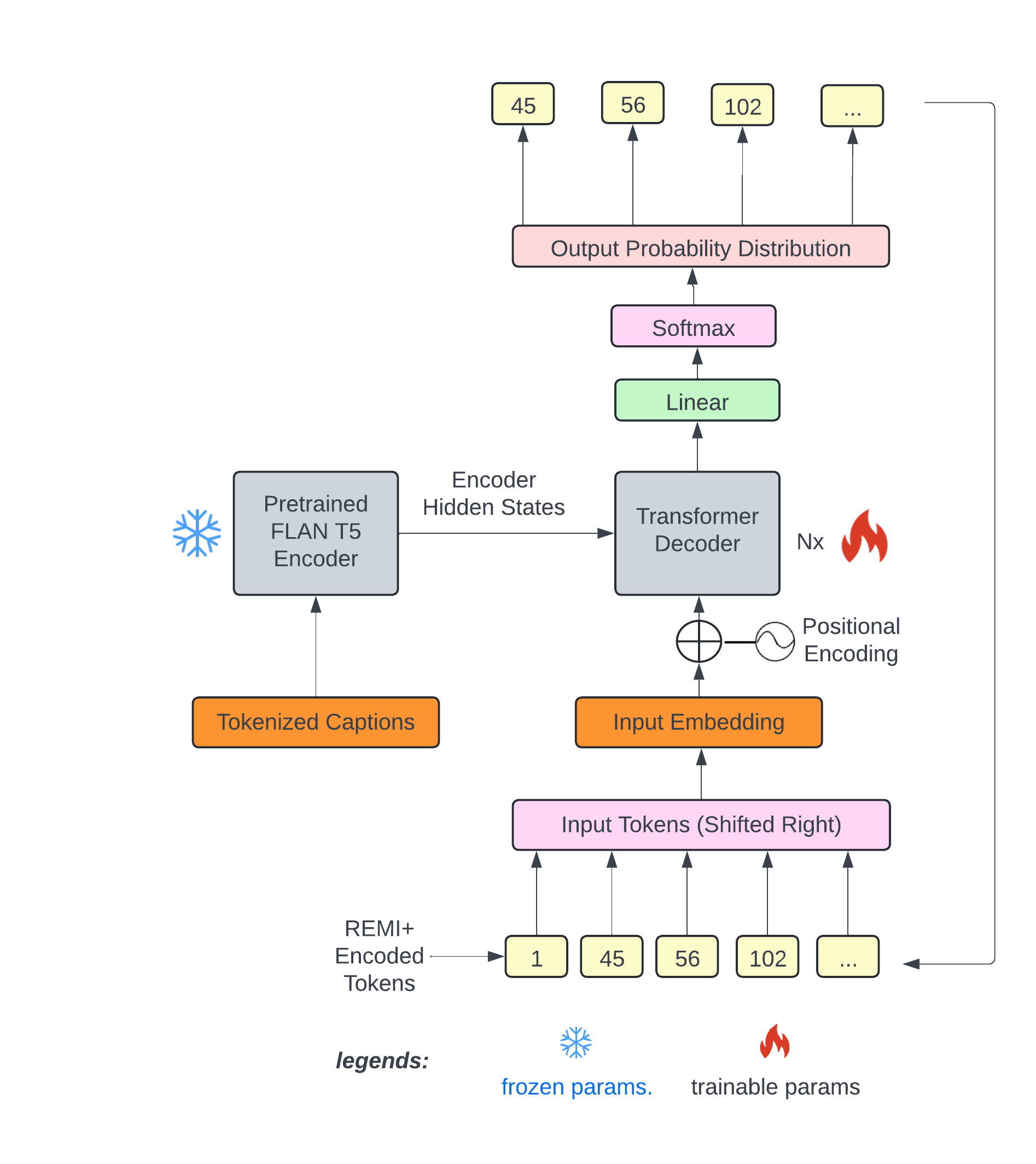}
\caption{Overview of our model's architecture. The FLAN T5 encoder weights are frozen. The transformer decoder accepts the encoder's hidden states via cross attention before generating the REMI+ encoded tokens autoregressively.}
\label{fig:model_architecture}
\end{figure}

As shown in Figure \ref{fig:model_architecture}, the proposed \model{} model consists of an encoder-decoder transformer architecture. We use the pretrained FLAN T5 model \cite{chung2024scaling} as our encoder to embed text captions before passing them to the Transformer \cite{vaswani2017attention} decoder layers with cross attention. Flan T5 advances the T5 model \cite{raffel2020exploring} through instruction fine-tuning on a diverse set of tasks. The instruction fine-tuning approach enables FLAN T5 to generalize better to unseen tasks, allowing it to perform tasks more effectively based on natural language instructions. As Flan T5 offers strong language understanding, evident from the zero shot and few shot learning benchmarks over regular T5, we postulate that its embeddings can effectively capture the semantic richness of text captions, translating textual descriptions into musical concepts accurately without re-training. Thus, we freeze the weights of the Flan T5 model during the training process.

\subsubsection{MIDI tokenizer}

We use the MidiTok library \cite{miditok2021} to encode multi-track multi-instrument MIDI files with the REMI+ tokenizer \cite{von2023figaro}, an extension of the REMI encoding \cite{huang2020pop}. The REMI tokenizer represents musical notes in a tokenized format, using tokens for tempo, pitch, velocity, duration, bar, onset, and positions within bars. This tokenizer is extended to REMI+ by adding MIDI program and time signature tokens, allowing it to handle multi-track and multi-instrument compositions with varying time signatures. 

\subsubsection{Transformer Decoder}

Given that we have represented the MIDI sequence as tokens, we can use a transformer decoder to generate these tokens autoregressively, much like a language generation task. We implemented the flash attention \cite{dao2022flashattention} algorithm in the Transformer Decoder, over the vanilla attention mechanism from the original study \cite{vaswani2017attention} so as to speed up the computation  as well as reduce the memory usage of the model, particularly for long sequence tasks such as this one. 

We utilize vanilla positional embeddings to encode the order of tokens in the sequence. After the final decoder layer, a linear projection layer is employed to map the decoder outputs to the vocabulary size. During training, we optimize the model using the categorical cross-entropy loss, defined as:

\begin{equation}
    \mathcal{L}_{\text{CCE}}(y, \hat{y}) = -\sum_{i=1}^{N} y_i \log(\hat{y}_i)
\end{equation}

where \(y_i\) is the true label and \(\hat{y}_i\) is the predicted probability for class \(i\).

\subsection{Semi-supervised pretraining}

The proposed \model{} model aims to generate music sequentially (from the beginning) based on textual prompts, which necessitates using the first $n$ encoded tokens per MIDI file rather than randomly cropping from the middle. This approach, however, limits our training data as we cannot utilize information beyond the model's context window. To address this limitation, we implement a two-stage training process. First, we pretrain on a larger dataset using placeholder text captions. For this pretraining dataset, which lacks predefined captions, we extract objective attributes from the MIDI files using the Music21 library\cite{cuthbert2010music21}. These attributes are: time signature, key signature, beats per minute (BPM), and instruments. We construct 10 different placeholder sentences incorporating these attributes, which are then processed by the Flan T5 encoder. An example of a placeholder caption is \textit{``Played at \textbf{114} beats per minute in \textbf{4/4} time signature and the key of \textbf{G\# minor}, classical piece with the following instruments: \textbf{clarinet, English horn, flute, horn, piccolo, trombone, and trumpet}.''}, where the text in bold are the placeholders. 

Following the pretraining phase, the model is fine-tuned on a dataset with more detailed captions that include both the objective attributes used during pretraining and subjective elements such as style and mood. This allows the model to learn sophisticated text-to-music mappings. During fine-tuning, a sentence omission technique is applied at each iteration, where 20\%-50\% of sentences are randomly omitted with a 50\% probability. This technique regularizes the model by exposing it to a more varied and unpredictable training set, preventing overfitting and enhancing its generalization to unseen data. Additionally, it serves as implicit data augmentation by creating a larger and more diverse set of training examples, helping the model learn to fill in gaps and generate coherent outputs from incomplete inputs. Finally, it prepares the model for real-world scenarios where users may provide minimal or incomplete captions, enhancing its practical effectiveness.


\section{Experimental Setup}
Evaluating generative models is a challenging task \cite{agres2016evaluation}. In order to establish the effectiveness of \model{}, we conduct both an objective evaluation of the results as well as a subjective evaluation in the form of a listening test. This allows us to thoroughly test both the resulting musical quality as well as how much the generated MIDI adheres to the input caption. In this section, we first detail the datasets and model configurationused to train the model in the experiments, followed by experimental configuration and evaluation metrics. Finally, we discuss results and findings of our experiments.

\subsection{Training datasets}
We work with two datasets during our training process: SymphonyNet for semi-supervised pretraining and MidiCaps for finetuning towards MIDI generation from captions. 
\paragraph{MidiCaps} is a dataset of 168,401 unique MIDI files with text captions \cite{melechovsky2024midicaps}. The MIDI files were originally provided in the Lakh MIDI dataset \cite{raffel2016learning} released under the CC-BY 4.0 license. Each MIDI file is paired with a music feature-rich caption. An example caption in the dataset: ``A melodic and happy pop song with a Christmas vibe, featuring piano, clean electric guitar, acoustic guitar, and overdriven guitar. The song is in the key of A major with a 4/4 time signature and a moderate tempo. The chord progression revolves around D, E6, D, and E, creating a motivational and loving atmosphere throughout the piece.''  We use the provided training set (~90\% of the data) to train the model in our experiments.

\paragraph{SymphonyNet} \cite{liu2022symphony} is a comprehensive dataset of symphonic music. It comprises 46,359 multi-instrument, multi-track MIDI files, predominantly symphonic, with an average duration of 4.26 minutes per file. We selected SymphonyNet for semi-supervised pretraining due to its high-quality, multi-track data, which aligns well with our model's requirements. The SymphonyNet MIDI files were augmented with pseudo-captions as described in the methods section. 

\subsection{Model Configuration}
We use an encoder-decoder transformer architecture for pretraining on SymphonyNet and finetuning on MidiCaps. As mentioned earlier, our encoder is a pretrained FLAN T5 model \cite{chung2024scaling}. Our transformer decoder consists of 18 layers and 8 attention heads, with 272M parameters amongst which 159M are trainable parameters. We observe that the MIDI tokenizer contains on average 4-5 minutes length of data per 2,000 tokens. Since the average length of tracks in SymphonyNet as well as MidiCaps is 4-5 minutes, we configured the input to the transformer decoder to be 2,048 MIDI tokens. Any smaller sequences in dataset are padded. For pretraining, we train for 100 epochs, with a batch size of 4 and gradient accumulation set to 4. For finetuning on MidiCaps, we trained for 30 epochs. For both runs, we use the Adam optimizer \cite{kingma2014adam} coupled with a cosine learning rate schedule with a warm-up of 20,000 steps. For pretraining, our base learning rate is \(1e^{-4}\) whereas for finetuning, we use a reduced base learning rate of \(1e^{-6}\). Our models are trained on 6 NVIDIA L40S 48 GB GPUs. 





\begin{table*}[http!] 
    \centering
    \begin{tabular}{l|c|cc}
    \toprule
        \textbf{Generated by:} & \textbf{MidiCaps} & \textbf{\model{}} & \textbf{MuseCoco} \\
        \midrule
        Question & \multicolumn{3}{c}{Avg. rating (1-7)} \\ 
         \midrule
        Musical Quality & 5.79 & 4.62 & 4.40 \\  
        Overall matching & 5.42 & 4.67 & 4.07 \\ 
        Genre matching & 5.54 & 4.98 & 4.40 \\  
        Mood matching & 5.70 & 5.00 & 4.32 \\  
        Key matching & 4.61 & 3.64 & 3.36 \\  
        Chord matching & 3.20 & 2.50 & 2.00 \\  
        Tempo matching & 5.89 & 5.42 & 4.94 \\  
         \bottomrule
    \end{tabular}
    \caption{Results of the listening study. Each question is rated on a Likert scale from 1 (very bad) to 7 (very good). The table shows the average ratings per question for each set of generated music.}
    \label{tab:results_general}
    \label{tab:results_expert}
\end{table*}

\subsection{Test datasets and baselines}
In order to maintain fairness with subjective results (based only on five generated examples), we consider 100 (5\%) randomly selected samples from the MidiCaps test set \cite{melechovsky2024midicaps}. We compare the MIDI generated by \model{} to the ground truth MIDI file from the MidiCaps dataset, as well as with MIDI generated by the MuseCoco model (xlarge) \cite{lu2023musecocogeneratingsymbolicmusic}\footnote{For detailed and latest results, readers are invited to visit our demo page \url{https://github.com/AMAAI-Lab/Text2midi}}.

\subsection{Objective evaluation metrics}
In the objective evaluation of our model, we address three questions - how much long-term structure and patterns the music contains, how similar the input text and generated music are and how close some of the important musical features are to that of the ground truth. To answer these, we use three types of  evaluation metrics as discussed below: 

\paragraph{Compression ratio} uses the COSIATEC algorithm \cite{meredith2013cosiatec} as used by \cite{chuan2018modeling} to measure the long-term structure and repeating patterns in music. We measure the compression ratio MIDI files to quantify amount of long-term structure in them.

\paragraph{CLAP} or Contrastive Language-Audio Pretraining \cite{laionclap2023} enables the extraction of joint latent representations of audio and text samples. We use an improved CLAP checkpoint (LAION CLAP\footnote{https://huggingface.co/laion/larger\_clap\_music\_and\_speech}), specifically trained on music and speech. We extract the latent representations of an input text (caption) and the synthesized audio of the MIDI file. We then use the cosine similarity between these latent representations as a measure of similarity between text (caption) and MIDI (synthesized audio). This enables us to evaluate how closely MIDI files fit their text caption.

\paragraph{Feature wise comparison} consists of four features extracted from generated MIDI files and their related ground truth MIDI files as used by \citet{melechovsky2023mustango}: 
\begin{itemize}
\item Tempo Bin (TB): the ratio of files of which the extracted tempo (in terms of beats per minute (bpm)), falls within the tempo bin extracted from the ground truth MIDI file, with bin borders defined at 40, 60, 70, 90, 110, 140, 160, 210 bpm. The tempo is extracted with music21 \cite{cuthbert2010music21}. 
%
\item Tempo Bin with Tolerance (TBT): this metric allows for slightly more tolerance than the previous one. It is calculated as the ratio of MIDI files of which the predicted bpm falls into the ground truth tempo bin or a neighboring one, compared to all files.
\item Correct Key (CK): the ratio of MIDI files of which the extracted key (using music21 \cite{cuthbert2010music21}) matches the extracted of the ground truth key MIDI file.

\item Correct Key with Duplicates (CKD): this metric allows for slightly more tolerance than the previous one. It is calculated as the ratio of generated MIDI files for which the extracted key matches the key of the extracted key of the ground truth MIDI key or an equivalent key (i.e., a major key and its relative minor).
\end{itemize}








\subsection{Listening test}
For the subjective assessment of our model, we conducted a listening study utilizing PsyToolkit \cite{stoet2010psytoolkit}. Participants were invited to listen to 15 rendered MIDI files: 5 randomly selected captions from the MidiCaps test dataset with their accompanying ground truth MIDI, 5 generated by \model{} using the text captions of the above 5 MIDI files from MidiCaps as input, and 5 generated by MuseCoco using the same text captions. We did not cherry pick the generated samples. These 15 samples were randomly ordered for each participant. Listeners evaluated these tracks across six criteria:
\begin{itemize}
\item Musical quality of the music
\item Overall match between the music and the caption
\item The genre of the music matches the caption
\item The mood of the music matches the caption
\item The key of the music matches the caption
\item The chords in the caption are prominent in the music
\item The tempo of the music matches the caption
\end{itemize}

\begin{figure*}[h!]
\centering
\includegraphics[width=.8\textwidth]{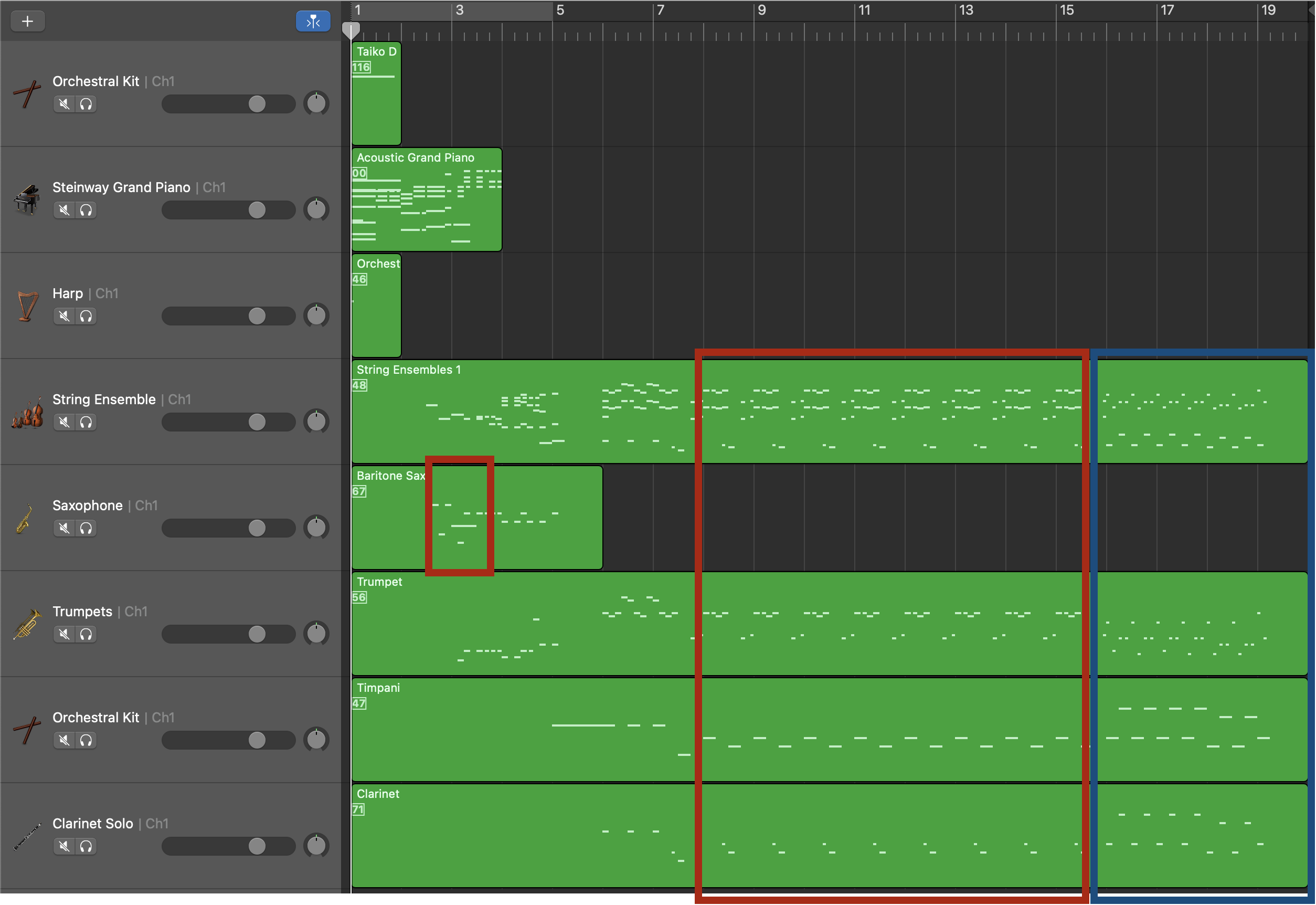}
\caption{A piano roll representation of a generated MIDI using the following caption ``A haunting electronic ambient piece that evokes a sense of darkness and space, perfect for a film soundtrack. The string ensemble, trumpet, piano, timpani, and synth pad weave together to create a meditative atmosphere. Set in F minor with a 4/4 time signature, the song progresses at an Andante tempo, with the chords F, Fdim, and F/C recurring throughout.''}
\label{fig:piano}
\end{figure*}

\section{Results and Discussion}
We assess various aspects with a particular focus on how the generated music reflects the captions, both in an objective experiment as well as a listening study. For all our evaluation tasks, we generate MIDI with 2,000 tokens. All our generated MIDI files typically last between 30-40 seconds when played back. For a fair comparison, while comparing with ground truth (gt) data, we only consider the first 40 seconds of the latter.

\subsection{Objective results}
ble \ref{tab:results} shows the results of the objective evaluation. Evaluation metrics are computed on MIDI files generated (from the same caption) by both \model{}, MuseCoco \cite{lu2023musecocogeneratingsymbolicmusic}, and the original MIDI files (ground truth) from the data set (MidiCaps). 

\begin{table}[h!]
    \centering
    \resizebox{\columnwidth}{!}{%
    \begin{tabular}{l|cccc}
    \toprule
      &  \textbf{Model} & \textbf{MidiCaps} & \textbf{MuseCoco} & \textbf{P-Val} \\
    \midrule
    CR $\uparrow$ & 2.31 & 3.43 & 2.12 & $<0.0001$ \\
    CLAP $\uparrow$ &  0.22 & 0.26 & 0.21 & 0.0102 \\
    \midrule
    TB (\%) $\uparrow$ & 39.70 & - & 21.71 & 0.1102\\
     TBT (\%) $\uparrow$    & 65.80 & - & 54.63 & 0.2051 \\
     CK (\%) $\uparrow$ & 33.60 & -  & 13.70 & $<0.0001$\\
     CKD (\%) $\uparrow$ & 35.60 & - & 14.59 & $<0.0001$\\
     \bottomrule
    \end{tabular}%
    } 
    \caption{Summary of our objective evaluations. Numbers are average results for all captions from the MidiCaps test set. CR: Compression ratio; CLAP: CLAP score; TB: Tempo Bin; TBT: Tempo Bin with Tolerance; CK: Correct Key; CKD: Correct Key with Duplicates; $\uparrow$: higher score better.}
    \label{tab:results}
\end{table}

Generating music with long-term structure is a known challenge \cite{bhandari2024motifs}. We use the compression ratio, as done by \citet{chuan2018modeling}, to evaluate the presence of repeated patterns in the MIDI files. \model{} achieves an average compression ratio of 2.31, significantly higher than MuseCoco's 2.12 ($p < 0.0001$). This result demonstrates that \model{} outperforms MuseCoco in creating long-term structure and repeating patterns. 

To evaluate the match between text caption and MIDI, we calculate the CLAP score. In Table \ref{tab:results_general}, we see that \model{}'s CLAP score (0.22) beats that of MuseCoco (0.21), which is quite encouraging. However, both models fall below the ground truth CLAP score (0.26), showing room for further improvement. We observed that, for the specific examples where the similarity between the text and ground truth MIDI is high (i.e. high CLAP score), the CLAP score for generated MIDI is also high. We thus posit that our model is able to adhere to the prompts and achieve better text-music alignment compared to our baseline.

While the CLAP score provides a general indication of how well the text and MIDI align, we further analyze specific aspects in which the generated MIDI resembled the ground truth MIDI. It is important to note that as these metrics compare the generated MIDI with the ground truth MIDI, they cannot be computed for MidiCaps, which serves as the ground truth here. We show that in terms of both tempo and key attributes, \model{} outperforms the MuseCoco model. The tempo bin (TB) metric reaches 39.70 compared to 21.71 for MuseCoco, and the TBT reaches 65.80 compared to 54.63 for MuseCoco. These ratios indicate that about 66\% of the tempo instructions are very close to the desired tempo. The CKD ratios for \model{} are also higher than MuseCoco (35.60\% vs 14.59\%), indicating better key matching. We note that both CK and CKD are statistically significant ($p<0.0001$). In terms of inference speed, we notice that \model{} is much faster compared to MuseCoco as a result of unifying the text-to-attribute and attribute-to-music models into a single holistic approach. Based on our observation, \model{} takes around 55 seconds to generate a 40-second long MIDI file, compared to 120 seconds for MuseCoco (inference performed on GPU in both cases). This comprehensive evaluation demonstrates that \model{} consistently surpasses MuseCoco on all objective metrics, despite training on a data set of only one fifth of the size used to train MuseCoco.



\subsection{Subjective results}

A total of 11 participants participated in the listening test. 3 individuals reported being able to recognize chords and keys with absolute pitch capabilities, while 6 individuals reported receiving more than one year of musical training experience.

The results of our subjective test are shown in Table \ref{tab:results_expert}. In general, all ratings for the ground truth MidiCaps examples gravitate towards `good' (score 5) and that of \model{} and MuseCoco generated examples, towards `neutral'(4) to `good' (5). On questions where the MidiCaps files receive lower scores, the generated examples also tend to score lower. The similar lower scores indicate a close alignment between the trained model and the original MidiCaps data. On average, although for all evaluation criteria MidiCaps examples fare better, they don't outperform \model{} generated examples by a large margin. \model{} generated examples also scored better in all questions compared to MuseCoco generated examples. This shows the effectiveness of \model{}'s ability to better follow the musicality of the MidiCaps ground truth. We have kept the listening test open for participation. Readers are requested to visit our GitHub page \footnote{\url{https://github.com/AMAAI-Lab/Text2midi}} for updated results.  

\subsection{Discussion}

The \model{} model exhibits strong capabilities in generating music with long-term structure and repeating patterns that make compositional sense, while still being able to break out of those repetitions, and to then make use of the same motifs from earlier in the composition. Figure \ref{fig:piano} shows the piano roll representation of a generated piece. An example of a repeated motif is shown in the red boxes, which first occurs at 3 seconds in, being used again with variations at 14 seconds, before switching to a different repeating pattern with a similar theme at 33 seconds (highlighted in blue). Such translated motifs would be recognized as patterns with COSIATEC \cite{meredith2013cosiatec} and hence leads to a higher compression ratio. The example also nicely shown that appropriate instrument tracks are generated to match the example (classical piece). The notes also appear `long' on the piano roll, indicating a match with the desired `slow and emotional' character. \model{} still struggles with instrumentation, which is apparent in this example: where a `classical' piece is asked for, the instruments definitely match this (e.g. trombone, french horn), however, the additional instruction of `church organ' is ignored. This could be due to the fact that the captions describe the entire song, whereas we only used part of the song for training, hence some instruments may have not started played. In future work, it would be interesting to explicitly focus on encoding instrumentation. 


\section{Conclusion}

In this paper, we introduce \model{}, the first end-to-end model for generating MIDI files directly from textual descriptions. The \model{} model leverages the power of pretrained large language models (LLMs) and combines it with an autoregressive transformer decoder to generate expressive MIDI files. The model is trained using semi-supervised learning on large-scale datasets including MidiCaps and SymphonyNet. 
In a detailed analysis, our analytical experiment shows that \model{} is able to generate MIDI pieces that have a long-term structure with repeating patterns, and show that the requested features in the input text caption are indeed adhered to in the generated MIDI. 

This work not only advances the state-of-the-art in text-to-MIDI generation but also opens up new avenues for intuitive music composition, by both making it accessible to a broader audience through the use of simple text prompts, as well providing a user-friendly tool for expert composers and producers to generate musical ideas. In future, we believe this model may be further trained to increase its output quality as well as be expanded to include more complex tasks including midi editing guided by text prompts. Another current limitation is the data quality, with higher quality creative commons MIDI files, model quality would further increase.


\section{Acknowledgments}
This work was supported by the UKRI and EPSRC under grant EP/S022694/1 and SUTD's Kickstart Initiative under grant number SKI 2021\_04\_06. We owe much appreciation to our reviewers for their insightful critiques which greatly strengthened the work.

\bibliography{main}

\end{document}